\documentclass[
 reprint,
 amsmath,amssymb,
 aps,
]{revtex4-2}

\usepackage{graphicx}
\usepackage{bm}

\usepackage[colorlinks=true, linkcolor=blue,anchorcolor=blue, citecolor=blue,urlcolor=blue]{hyperref}

\usepackage[normalem]{ulem}
\begin{document}
\preprint{APS/123-QED}


\title{Strain-Induced Intrinsic Antiferromagnetic Skyrmions in Two-Dimensional Janus Magnets}

\author{Weiyi Pan$^{1}$}
\email{pwy20@mails.tsinghua.edu.cn}
\author{Zhiming Xu$^{1}$}

\affiliation{$^{1}$State Key Laboratory of Low Dimensional Quantum Physics and Department of Physics, Tsinghua University, Beijing, 100084, China\\}

\date{2024.3}

\begin{abstract}
Antiferromagnetic (AFM) skyrmions, which are resistant to both the skyrmion Hall effect and external magnetic perturbations, are expected to be promising candidates for next-generation spintronics devices. Despite being observed in bulk materials and synthetic AFM layered systems, the existence of intrinsic AFM skyrmions within single magnetic layers, which offer potential advantages for spintronic device fabrication, has remained elusive. In this work, taking monolayer CrSi(Te,Se)$_{3}$ as a representative system, we demonstrate the emergence of intrinsic AFM skyrmions in two-dimensional Janus magnets. It is found that under moderate compressive strain, the interplay between considerable Dyzaloshinskii-Moriya interaction and the strain-induced AFM Heisenberg exchange interaction in monolayer CrSi(Te,Se)$_{3}$ would give rise to the emergence of intrinsic AFM skyrmions assembled from AFM spin spirals. Moreover, the application of an external magnetic field could trigger the emergence of AFM merons as well as a canted AFM state. Our findings propose a feasible approach for achieving intrinsic AFM skyrmions in realistic systems, which paves the way for developments in AFM topological spintronics devices.

\end{abstract}
\maketitle
\section{INTRODUCTION}

Magnetic skyrmions \cite{skyrmion1,skyrmion2,skyrmion3}, which are topologically protected nanosized spin textures with unique physical properties \cite{SkxME1,SkxME2,SkxME3,SkxME4} and promising applications \cite{SkXApp1,SkXApp2,SkXApp4}, have received great attention in recent years. To date, the majority of skyrmions are known as ferromagnetic (FM) skyrmions, since they have been observed in a FM background. However, FM skyrmions exhibit two remarkable drawbacks. First, their nonzero topological charge leads to the skyrmion Hall effect \cite{SkHE}, causing deviations in their trajectory when they are driven by external currents, thereby hindering precise control over their motion. Second, FM skyrmions are susceptible to interactions with external magnetic perturbations \cite{SkHE,Defect,Defect2}, rendering them insufficiently robust for future applications.

In contrast, antiferromagnetic (AFM) skyrmions offer distinct advantages due to their resistance to the skyrmion Hall effect and insensitivity to external magnetic perturbations \cite{AFM,AFM2,AFM3,AFM4,AFM5,Samir}. These intrinsic AFM topological textures have been extensively studied in bulk systems \cite{bulk1,bulk2}. Compared to bulk materials, the inherent thinness of two-dimensional (2D) materials provides enhanced integration capabilities for electronic devices \cite{Nano,2D,2D2}, and thus there is considerable anticipation about AFM skyrmions in 2D magnets. One effective strategy for realizing AFM skyrmions in 2D systems is to construct synthetic AFM skyrmions \cite{Syn,Syn-AFM,Syn-AFM2,Nano1}. Each synthetic AFM skyrmion consists of two FM skyrmions positioned in separate magnetic layers, with the layers coupled antiferromagnetically. However, achieving such synthetic AFM skyrmionic systems requires precise structural engineering to adjust the interlayer coupling to an AFM-like state, which complicates device fabrication \cite{Syn-AFM,Syn-AFM2,Nano1}.

Alternatively, intrinsic AFM skyrmions situated within a single magnetic layer \cite{A1,A2,A3,A4,A5} have been conceptually proposed to address the aforementioned challenges. Two critical factors are essential for realizing the intrinsic AFM skyrmions: first, a significant Dyzaloshinskii-Moriya interaction (DMI) must exist within the selected single 2D layer; second, the intralayer exchange interaction should be AFM-like. In this scenario, the interplay between the AFM exchange interaction and the DMI would give rise to the emergence of intrinsic AFM skyrmions in a single layer. Given these considerations, a suitable system for realizing the intrinsic AFM skyrmion is Janus magnetic system, in which a substantial DMI may arise due to broken inversion symmetry \cite{Janus1,Janus2,Zhong,Yang2,Janus3}. Moreover, it is feasible for the AFM exchange couplings in 2D Janus magnetic systems to be achieved by strain. This arises from the sensitivity of the competition between AFM direct exchange and FM indirect exchange, which determines the sign of exchange coupling, to the lattice structure of 2D magnets \cite{Zhong,Yang2}. However, the achievement of intrinsic AFM skyrmions in realistic 2D Janus systems remains elusive.

In this study, we demonstrate that intrinsic AFM skyrmions could be achieved in Janus CrSi(Te,Se)$_{3}$ monolayer, which can be achieved by substituting Te atoms on the top layer of monolayer CrSiTe$_{3}$\cite{U2} with Se atoms. Through first-principles calculations and Monte Carlo (MC) simulations, we find that under moderate compressive strain, the considerable DMI could compete with the strain-induced AFM exchange interaction, which leads to the emergence of intrinsic AFM skyrmions assembled from AFM spin spiral (SS) states in monolayer CrSi(Te,Se)$_{3}$. Furthermore, magnetic field could induce AFM merons as well as a canted AFM state in the present system. Our results thus offer a promising path toward 2D AFM spintronic devices.

\section{METHODS}
\subsection{First-principles calculations}
First-principles calculations based on the projector augmented-wave (PAW) method were carried out using the Vienna \textit{ab} \textit{initio} simulation package (VASP)\cite{VASP,PAW}. During the calculation, we set the plane wave energy cutoff to 500 eV, and the calculation was performed with a $12\times12\times1$ Gamma-centered k-point mesh for the unit cell. The exchange correlation effect was included with the Perdew-Burke-Ernzerhof (PBE) functional under the generalized gradient approximation (GGA)\cite{PBE}. To describe the onsite electronic correlation effect of d orbitals for Cr, the Hubbard \textit{U} of 0.5 eV is used\cite{U}, which has been shown to be accurate for describing the magnetism of the CrXTe$_{3}$ (X = Ge, Si) family \cite{U6,U7,U2,U5,U8}. 
A 20 \AA ~ vacuum layer is adopted for all calculations. The convergence criteria for the total force and energy were set as 10$^{-3}$ eV/\AA ~ and 10$^{-6}$ eV, respectively. The phonon dispersion is calculated using the PHONOPY code\cite{Phonopy}. The molecular dynamics (MD) simulations are performed in the canonical (NVT) ensemble with a Nos$\Acute{\textup{e}}$ thermostat, where a $3\times3\times1$ supercell is adopted. The simulation processes 5000 steps at 1 fs per step. 

\begin{figure*}[ht]
\includegraphics[scale = 0.32 ]{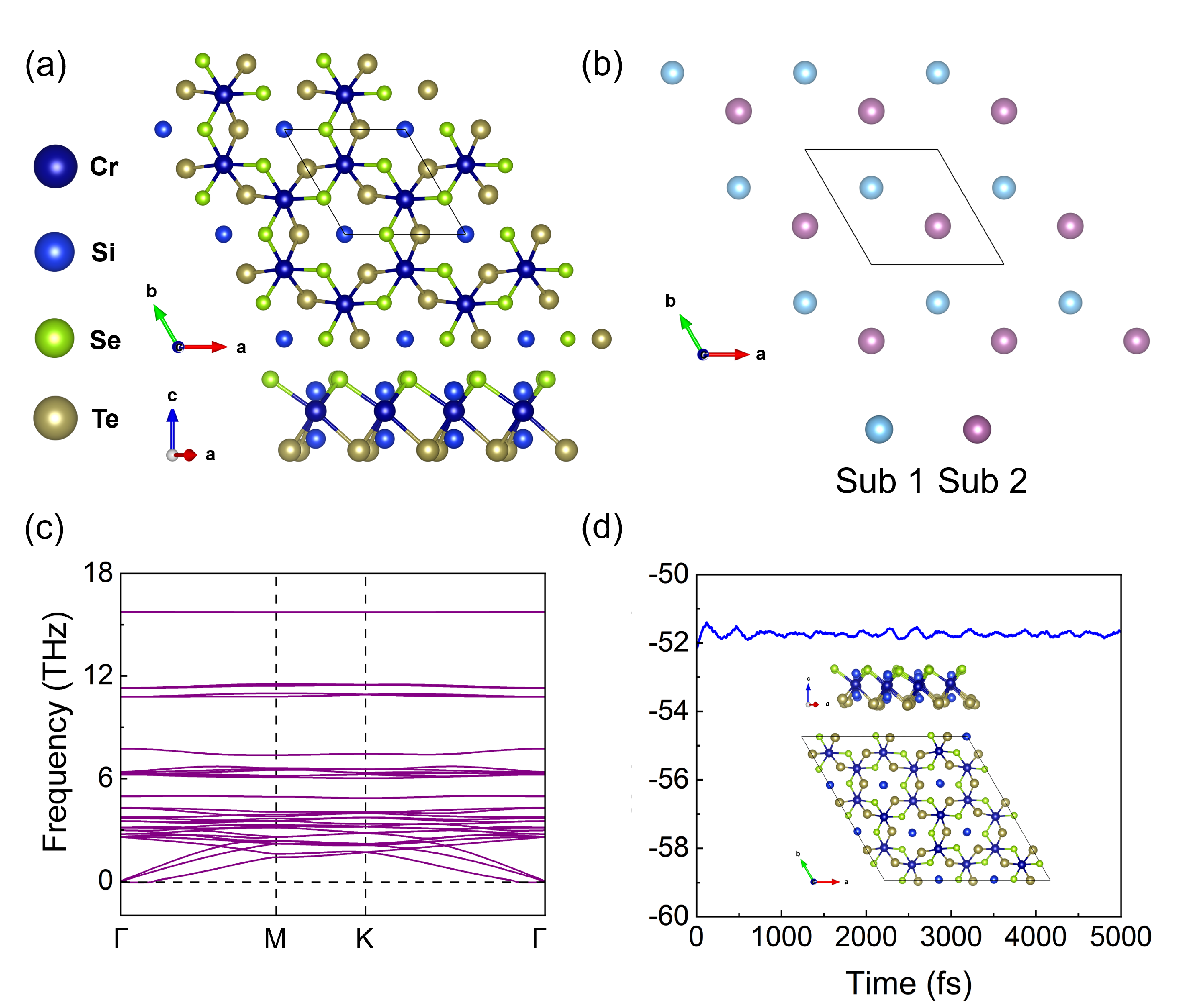}
\caption{\label{1} (a) Top and side views of the atomic structure of monolayer CrSi(Te,Se)$_{3}$. (b) Two nonequivalent magnetic atoms in a single unit cell, each of which makes up a triangular magnetic sublattice. (c) Phonon dispersion of monolayer CrSi(Te,Se)$_{3}$. (d) MD simulations of monolayer CrSi(Te,Se)$_{3}$ at 300 K. The evolution of energy is given,  while the inset shows the crystal structure after 5000 fs.  }
\end{figure*}

 \begin{figure*}[ht]
\includegraphics[scale = 0.44 ]{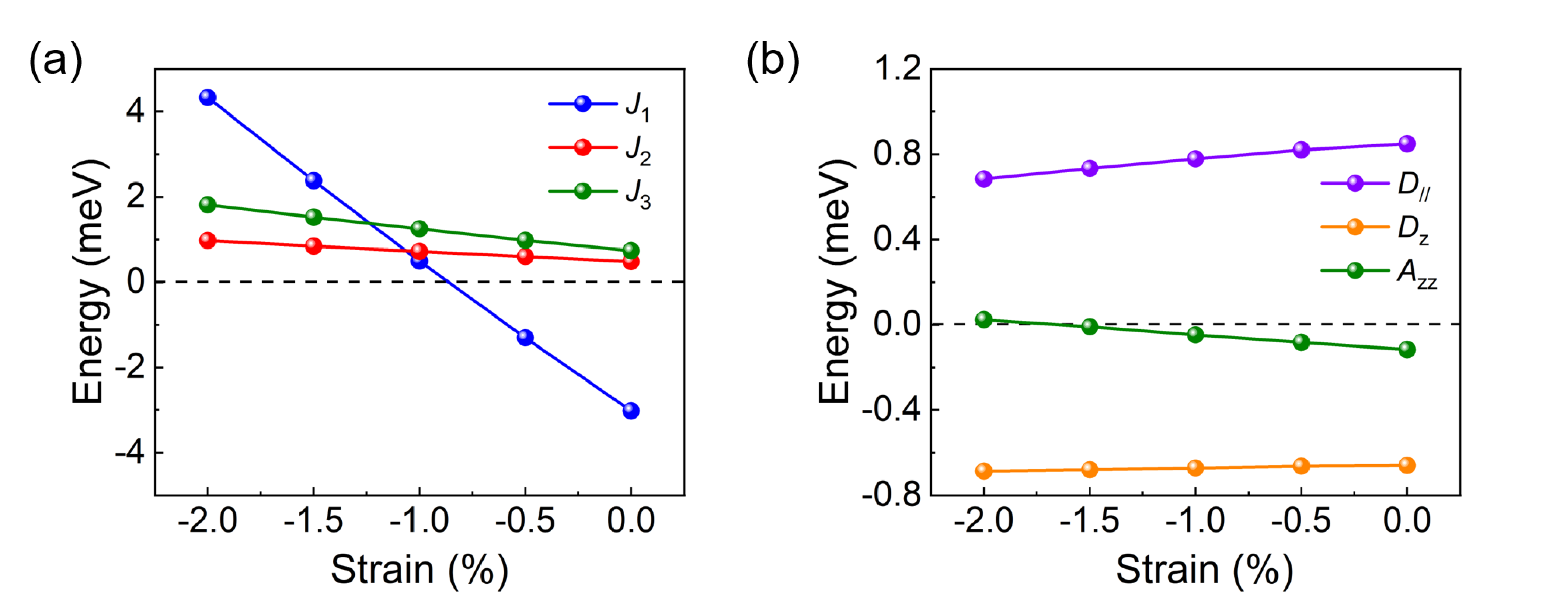}
\caption{\label{2} The dependence of magnetic parameters on compressive strain. (a) The value of the Heisenberg exchange as a function of compressive strain. (b) The values of the DMI and SIA as a function of compressive strain. Note that we renormalize the spin to S = 1 in our study for simplification. }
\end{figure*}

\subsection{Monte Carlo simulations}
Parallel tempering Monte Carlo (PTMC) simulations are carried out to obtain the classical magnetic configurations, as implemented in the PASP\cite{PASP1,PASP2} package. The MC simulations are performed on a $60 \times 30 \times 1$ supercell, which contains 7200 Cr atoms. Here, the supercell is based on the rectangular cell defined by
 $\textbf{a}' = \textbf{a}$, $\textbf{b}' = \textbf{a} + 2\textbf{b}$, and $\textbf{c}' = \textbf{c}$, 
 where $\textbf{a}$, $\textbf{b}$, and $\textbf{c}$ 
 are the lattice vectors of the original unit cell of the system. During the PTMC simulations, we set the initial spin configurations randomly, and 150 000 MC steps are performed for each configuration. Specifically, 500 exchange steps are set, while 300 MC steps are performed between two replica exchange processes. Moreover, 60 000 statistically independent samples are used during the tempering process. The initial temperature is set as 150 K, while the final temperature is 1 K. To further optimize the spin configurations obtained by MC simulation, we perform conjugate gradient (CG) optimization after the MC simulation, which guarantees that the resulting spin textures all lie at energy minima\cite{CG}. During CG optimization, the orientation of each spin is rotated until the force on each spin is minimized. The energy convergence criterion was set to 10$^{-6}$ eV.

\section{RESULTS AND DISCUSSION}
\subsection{Strain-induced magnetic transition}
The atomic structure of monolayer CrSi(Te,Se)$_{3}$ is depicted in Fig. \ref{1}(a). This structure comprises a top layer of Se atoms and a bottom layer of Te atoms, with both layers bridged by Cr atoms. This monolayer exhibits asymmetry between its top and bottom layers, without inversion symmetry, which is a typical characteristic of the Janus system. Therefore, it intrinsically possesses DMI between two adjacent Cr atoms. Notably, a unit cell contains two nonequivalent Cr atoms, each forming a triangular magnetic sublattice within an overall hexagonal magnetic lattice, as shown in Fig. \ref{1}(b).

\begin{figure*}[ht]
\includegraphics[scale = 0.44 ]{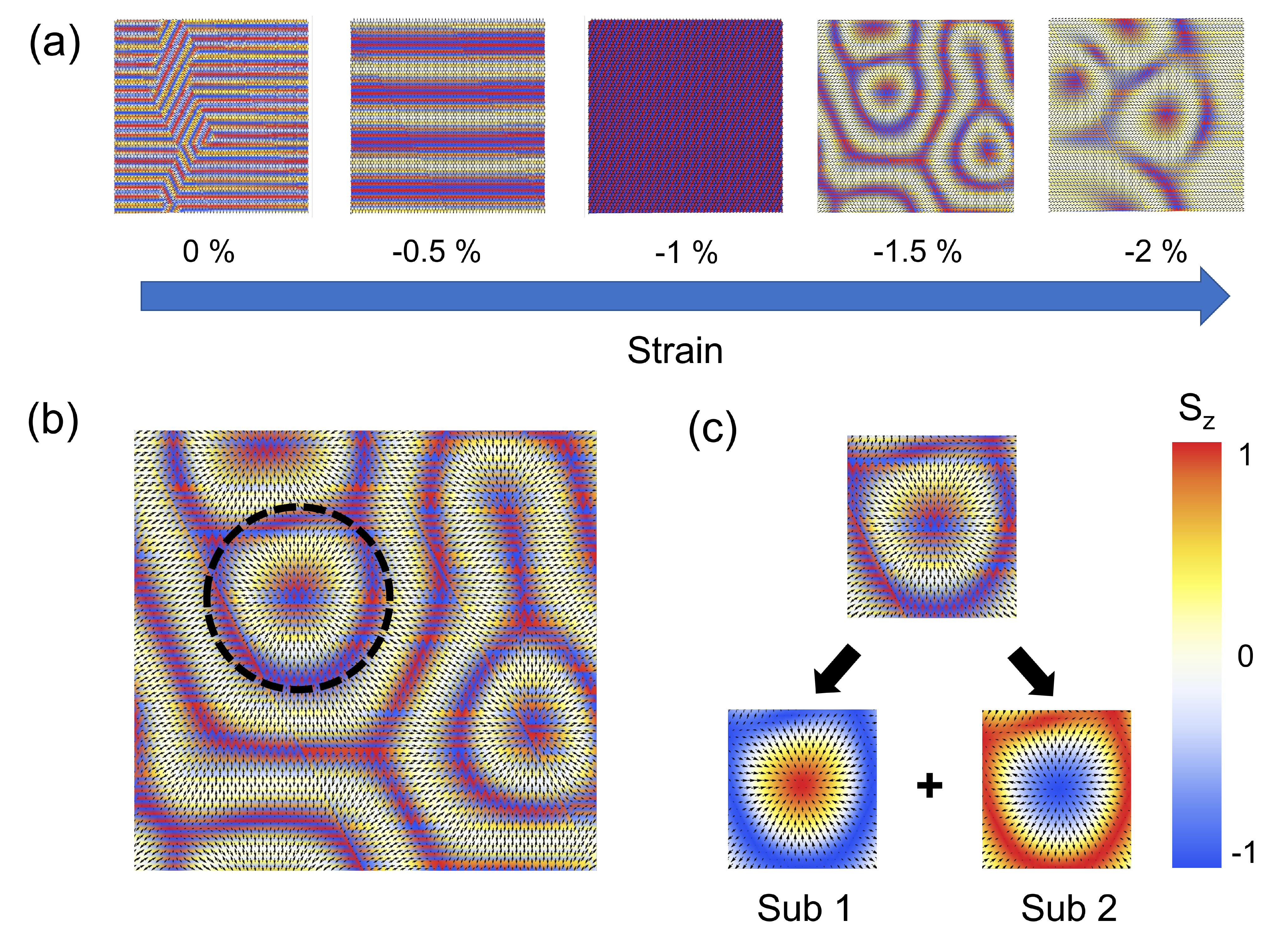}
\caption{\label{3} The dependence of magnetic textures on compressive strain. (a) Macroscopic magnetic states under distinct strain levels. (b) Close view of the magnetic texture under -1.5\% strain. A single isolated AFM skyrmion is circled, (c) which is composed of two FM skyrmions on each magnetic sublattice. }
\end{figure*}

We evaluate the formation energy $E_{f}$ of monolayer CrSi(Te,Se)$_{3}$, which is defined as $E_{f}$ = $E_{\textup{CrSi(Te,Se)}_{3}}$ -2$E_{\textup{Cr}}$-2$E_{\textup{Si}}$-3$E_{\textup{Te}}$-3$E_{\textup{Se}}$, where $E_{\textup{CrSi(Te,Se)}_{3}}$ is the total energy of monolayer CrSi(Te,Se)$_{3}$ per unit cell, and $E_{X}$ (X = Cr, Si, Te, Se) are the energies per atom for corresponding bulk phases of elemental X (X = Cr, Si, Te, Se). To be specific, the Cr atoms crystallize in a body-centered cubic phase\cite{Cr}, the Si atoms crystallize in a diamond structure\cite{Si}, and Se/Te atoms crystallize in trigonal phases\cite{SeTe,SeTe2}. The formation energy is calculated to be a negative value of -3.64 eV, which proposes that the monolayer CrSi(Te,Se)$_{3}$ is chemically stable. Moreover, by calculating the phonon dispersion of the system, no imaginary frequency is found, as shown in Fig. \ref{1}(c). This suggests that monolayer CrSi(Te,Se)$_{3}$ is dynamically stable. In addition, MD simulations are performed at 300 K, and the results can be seen in Fig. \ref{1} (d). The slight fluctuation of energy as well as the well-defined atomic structures of the system during the MD simulation confirm the thermal stability of monolayer CrSi(Te,Se)$_{3}$ at room temperature.

To reveal the magnetic interactions in monolayer CrSi(Te,Se)$_{3}$, we adopt the following spin Hamiltonian, which could describe the salient magnetic features of Janus magnets\cite{Zhong,Yang2}:
 \begin{equation}
 \begin{aligned}
     &   H = \sum_{\langle i,j \rangle} J_{1} \textbf{S}_{i} \cdot \textbf{S}_{j}  +  \sum_{\langle \langle i,j \rangle \rangle}J_{2}\textbf{S}_{i} \cdot \textbf{S}_{j} +  \sum_{\langle \langle \langle i,j \rangle \rangle \rangle}J_{3}\textbf{S}_{i} \cdot \textbf{S}_{j}  \\
     & +\sum_{\langle i,j \rangle} \textbf{D}_{ij} \cdot (\textbf{S}_{i} \times \textbf{S}_{j}) +\sum_{i}A_{zz}(S_{iz})^{2} 
 \end{aligned}
\end{equation}
in which $J_{1}$, $J_{2}$, and $J_{3}$ denote the first- (1NN), second- (2NN), and third-nearest-neighboring (3NN) Heisenberg exchange interactions, respectively. $\textbf{D}_{ij}$ represents the DMI between spins $S_{i}$ and $S_{j}$, and $A_{zz}$ is the single ion anisotropy energy (SIA). Structurally, a mirror plane exists perpendicular to the 1NN Cr-Cr bond in our system. Thus, the DMI vector between two 1NN Cr atoms can be expressed as $\textbf{D}_{ij} = D_{//}\hat{\textbf{z}}\times\hat{\textbf{u}}_{ij} + D_{z}\hat{\textbf{z}}$, where $\hat{\textbf{u}}_{ij}$ represents the unit vector pointing along the Cr-Cr bond and $\hat{\textbf{z}}$  represents the unit vector along the z direction. This expression of the DMI vector is commonly utilized for Janus systems with the C$_{3v}$ point group\cite{Zhong,Yang2}.

\begin{figure*}[ht]
\includegraphics[scale = 0.33 ]{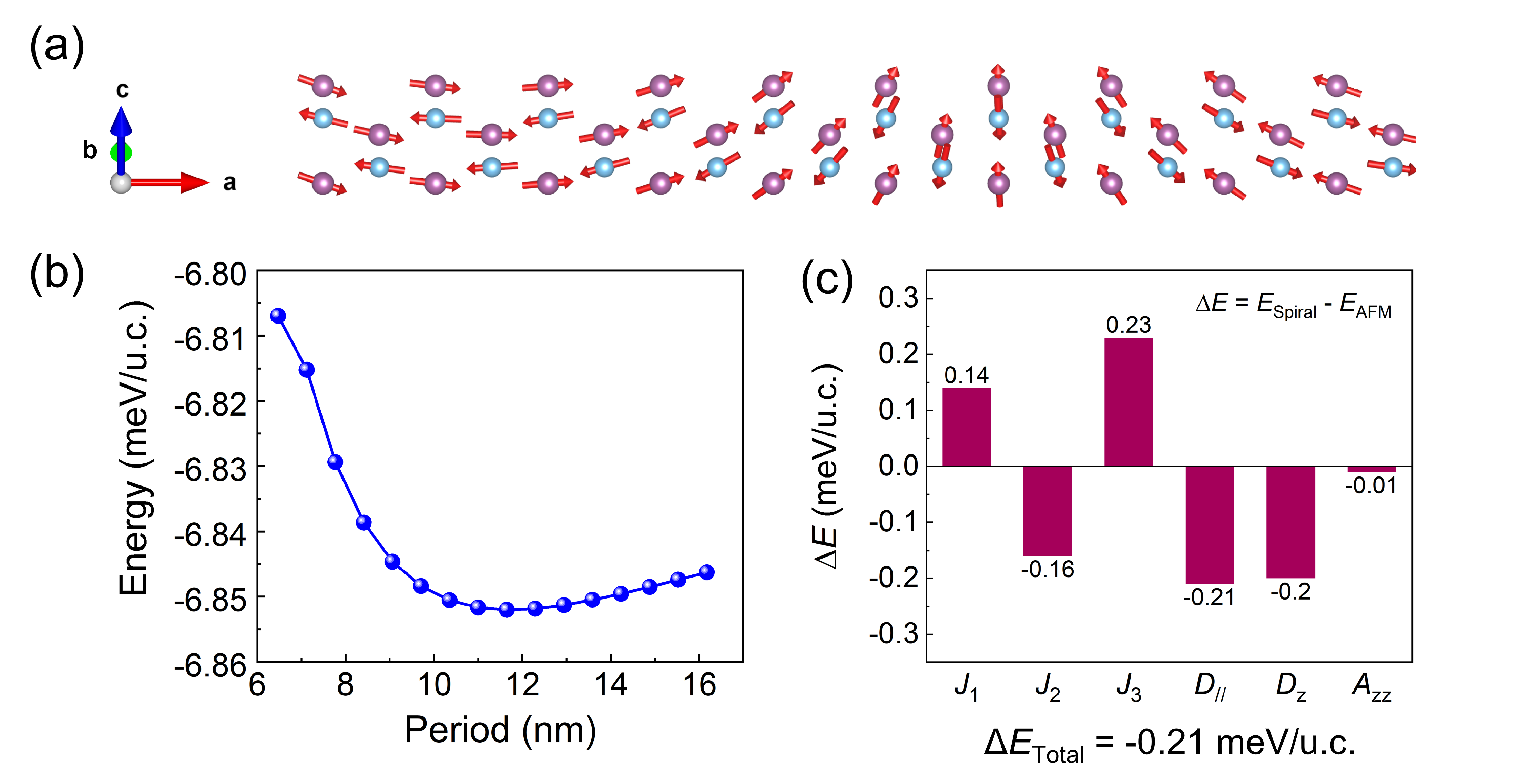}
\caption{\label{4} The magnetic ground state of monolayer CrSi(Te,Se)$_{3}$ under -1.5\% strain. (a) Schematic illustration of an AFM spin spiral (SS), which is the magnetic ground state of the system. Note that only half a period of SS is shown. (b) The energy of an AFM SS as a function of its period, in unit of meV per unit cell (u.c.). The lowest-energy period is 11.65 nm.(c) Decomposed energy difference between SS and collinear N$\Acute{\textup{e}}$el-type AFM, in unit of meV per unit cell (u.c.). The total energy difference is shown below.}
\end{figure*}

\begin{figure*}[ht]
\includegraphics[scale = 0.32 ]{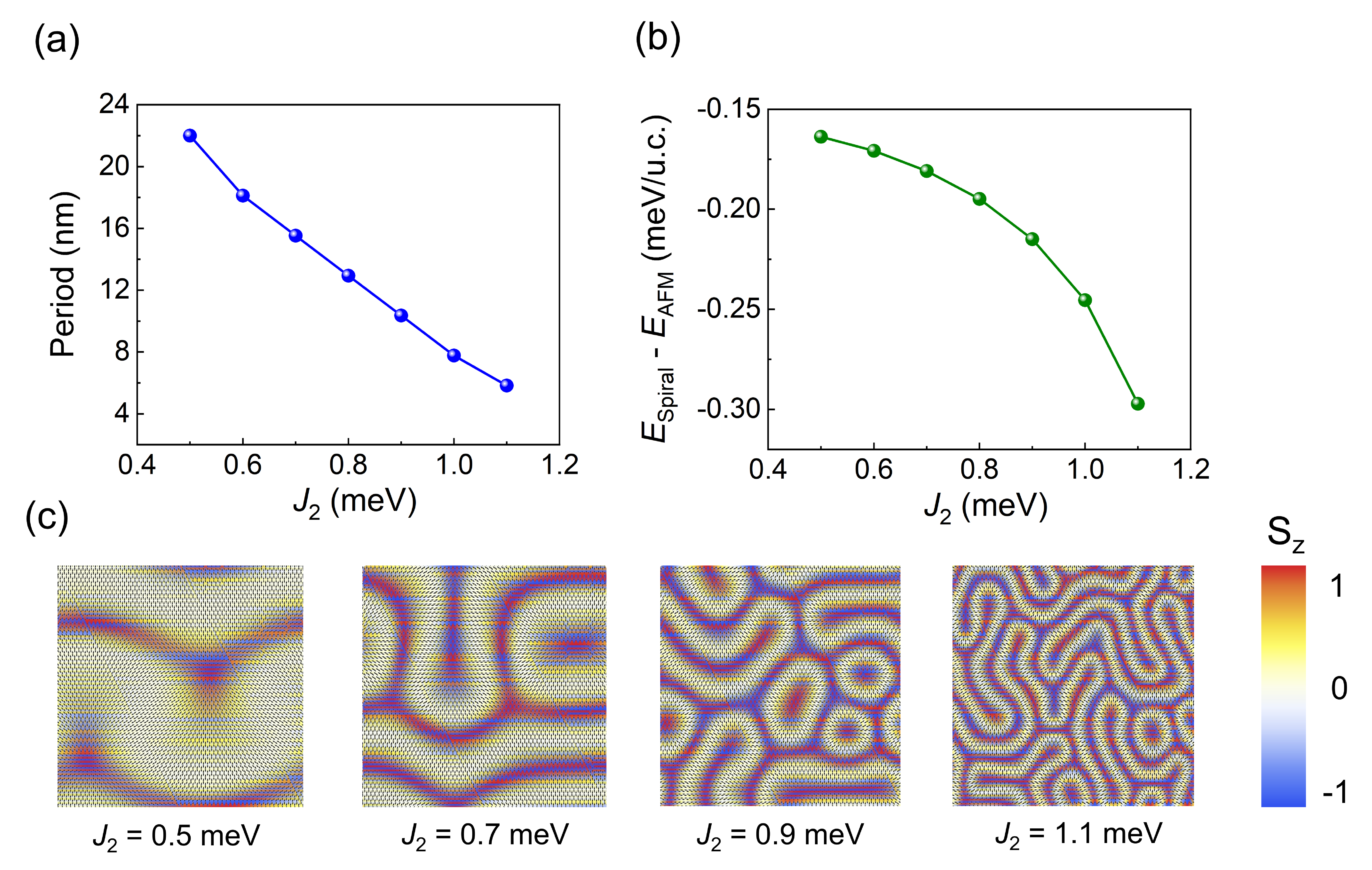}
\caption{\label{5} Effect of $J_{2}$ on the magnetic ground state. (a) Variation in the spin spiral period as a function of $J_{2}$. (b) Energy difference between the lowest-energy spin spiral and collinear N$\Acute{\textup{e}}$el-type AFM states as a function of $J_{2}$, in unit of meV per unit cell (u.c.). (c) Evolution of the macroscopic LD state under distinct values of $J_{2}$.}
\end{figure*}

By mapping the energies of distinct spin configurations onto equation (1), we derived the values of the magnetic parameters in the equation (see the supplementary material for more details\cite{SM}). For simplicity, we renormalize the spins to S = 1 in our study. As shown in Fig. \ref{2}, the pristine system without strain exhibits a significant DMI, and the 1NN Heisenberg exchange $J_{1}$ is FM. Consequently, $J_{1}$ ferromagnetically couples the spins on the two sublattices, inhibiting the emergence of AFM topological spin textures. In addition to the FM exchange between 1NN Cr atoms, there are smaller but nonnegligible AFM exchanges $J_{2}$ at 2NN and $J_{3}$ at 3NN. Moreover, the SIA is extremely small and has a negligible influence on the spin orientation of each Cr ion in the system. By applying external compressive strain to the system, the value of parameter $J_{1}$ increases monotonically, transitioning from a negative to a positive value. This transition corresponds to the switching from FM to AFM coupling between the two magnetic sublattices, as shown in Fig. \ref{1}(b). Microscopically, this strain-induced transition from FM to AFM in $J_{1}$ corresponds to an increase in direct AFM exchange between Cr atoms, which is attributed to the decrease in the Cr-Cr bond distance (see the SM for more details \cite{SM}). In addition to $J_{1}$, other magnetic parameters such as $J_{2}$, $J_{3}$, DMI, and SIA, exhibit only slight changes with strain. It is particularly significant that as the compressive strain reaches a moderate level of -1.5\%, the AFM $J_{1}$ becomes notably strong. Moreover, the ratio of $\mid D/J_{1}\mid $ reaches 0.42, a value notably higher than the typical range of 0.1 to 0.2 observed in skyrmionic systems \cite{Janus2}. Thus, it is reasonable to speculate that the interplay between substantial AFM $J_{1}$ and the considerable nonzero DMI in the strained monolayer CrSi(Te,Se)$_{3}$ makes it a promising candidate for hosting AFM topological spin textures. We emphasize that achieving a -1.5\% strain is experimentally viable through epitaxial growth \cite{tian2021manipulating,miao2021straintronics,cheng2024tunable}, and thus it is possible to practically realize the strain-induced FM-to-AFM transition in our system.

To gain deeper insights into the strain-dependent spin textures, we perform MC simulations on large supercells with magnetic parameters calculated under different strain levels. The resulting macroscopic magnetic textures are displayed in Fig. \ref{3}(a). Under compressive strains not exceeding -1\%, stripe-like spin textures are clearly observed. This result stems from the competition between FM coupling ($J_{1}$) and AFM coupling ($J_{2}$ and $J_{3}$). However, at a compressive strain of -1.5\%, a distinctive AFM labyrinth domain (LD) state emerges, characterized by various isolated skyrmions against an N$\Acute{\textup{e}}$el-type AFM background, as depicted in Fig. \ref{3}(b). Notably, each individual AFM skyrmion consists of two AFM-coupled FM skyrmions within each magnetic sublattice, as illustrated in Fig. \ref{3}(c) and Fig. S3 in the SM\cite{SM}. In contrast to the widely reported synthetic AFM skyrmions, which consist of two FM skyrmions located in separate magnetic layers with interlayer AFM couplings \cite{Syn-AFM,Syn-AFM2}, the AFM skyrmions intrinsically reside within a single atomic layer and do not require an external magnetic field. These zero-field intrinsic AFM skyrmions, which are located in a magnetic monolayer and have rarely been reported previously \cite{Samir}, not only provide a platform for investigating exotic physics associated with AFM skyrmions but also offer significant advantages for AFM spintronic devices. Upon further increasing the compressive strain to -2\%, both the size of the AFM LDs and skyrmions increase due to the increase in the value of $J_{1}$. This increase, however, results in a reduction in the storage density of skyrmion-based memory devices, which undermines their practical applicability. Therefore, we will focus on the case of -1.5\% strain in the following text.

\subsection{Microscopic origin of AFM spin spirals and skyrmions}

To gain a deeper understanding on the formation of AFM skyrmions in monolayer CrSi(Te,Se)$_{3}$, it is necessary to determine the nature of the magnetic ground state. To this end, we carried out MC simulations followed by CG optimization on supercells of different sizes, resulting in the identification of various low-energy AFM SS states against the N$\Acute{\textup{e}}$el-type AFM background and with distinct periods. The magnetic ground state of the system is determined to be an AFM SS state with a period of 11.65 nm, as depicted in Fig. \ref{4}(a) and \ref{4}(b). This AFM SS exhibits two primary characteristics: (1) it is composed of two FM chiral SSs winding in the x-z plane across each magnetic sublattice, and (2) the in-plane components of two nearest-neighboring spins are not strictly antiparallel; instead, they exhibit a canting angle of approximately 170° (see Fig. S4(a) in SM \cite{SM}). These AFM SSs serve as fundamental units that can assemble  a metastable LD state, within which various isolated AFM skyrmions can exist.

To clarify the mechanism stabilizing the SS state against the N$\Acute{\textup{e}}$el-type AFM background, we calculate the energy difference between the lowest-energy AFM SS state and the collinear N$\Acute{\textup{e}}$el-type AFM state (see Fig. S1 (a) in SM \cite{SM}) with out-of-plane magnetization. The energy of the lowest-energy AFM SS is 0.21 meV per unit cell lower than that of the collinear N$\Acute{\textup{e}}$el-type AFM state. Note that the microscopic LDs containing AFM skyrmions are identified as metastable states, as shown in Fig. \ref{3}(c), which are only 0.01 meV per unit cell higher in energy than that of the lowest-energy AFM SS state. As a result, metastable AFM skyrmions are achievable in our system. By further decomposing the energy difference into distinct magnetic parameters, we can determine the interactions that promote the emergence of the AFM SS state. As shown in Fig. \ref{4}(c), the DMI contributes most significantly to the energy gains of the AFM SS state. By turning off the DMI and rerunning the MC simulation, the SS collapses into a trivial out-of-plane collinear N$\Acute{\textup{e}}$el-type AFM state. This confirms that the DMI plays a predominant role in stabilizing the AFM SS as the ground state, which is vital for the emergence of metastable AFM skyrmions made up of AFM SSs in the system, consisting with previous reports \cite{Janus1,Janus2,Janus3}. To investigate the specific role of each DMI component, we switch $D_{//}$ and $D_{z}$ on and off in turn during the MC simulations, while other interactions are kept unchanged. By turning off $D_{z}$, the SS state still exists, yet the in-plane components of two nearest-neighboring spins do not cant with each other, but form a strict anti-parallel arrangement (see Fig. S4(b) in SM \cite{SM}). On the other hand, by switching off 
$D_{//}$, the SS collapses into an in-plane N$\Acute{\textup{e}}$el-type AFM state with canted spins (see Fig. S4(c) in SM \cite{SM}). 
Thus, we confirm that the in-plane DMI component $D_{//}$ would give rise to SS in which spins could rotate in the x-z plane, while the out-of-plane DMI component $D_{z}$ induces the in-plane canting between two nearest-neighbouring spins, which is consistent with previous knowledge for Janus systems with hexagonal lattices \cite{Janus2}.

\begin{figure*}[ht]
\includegraphics[scale = 0.34 ]{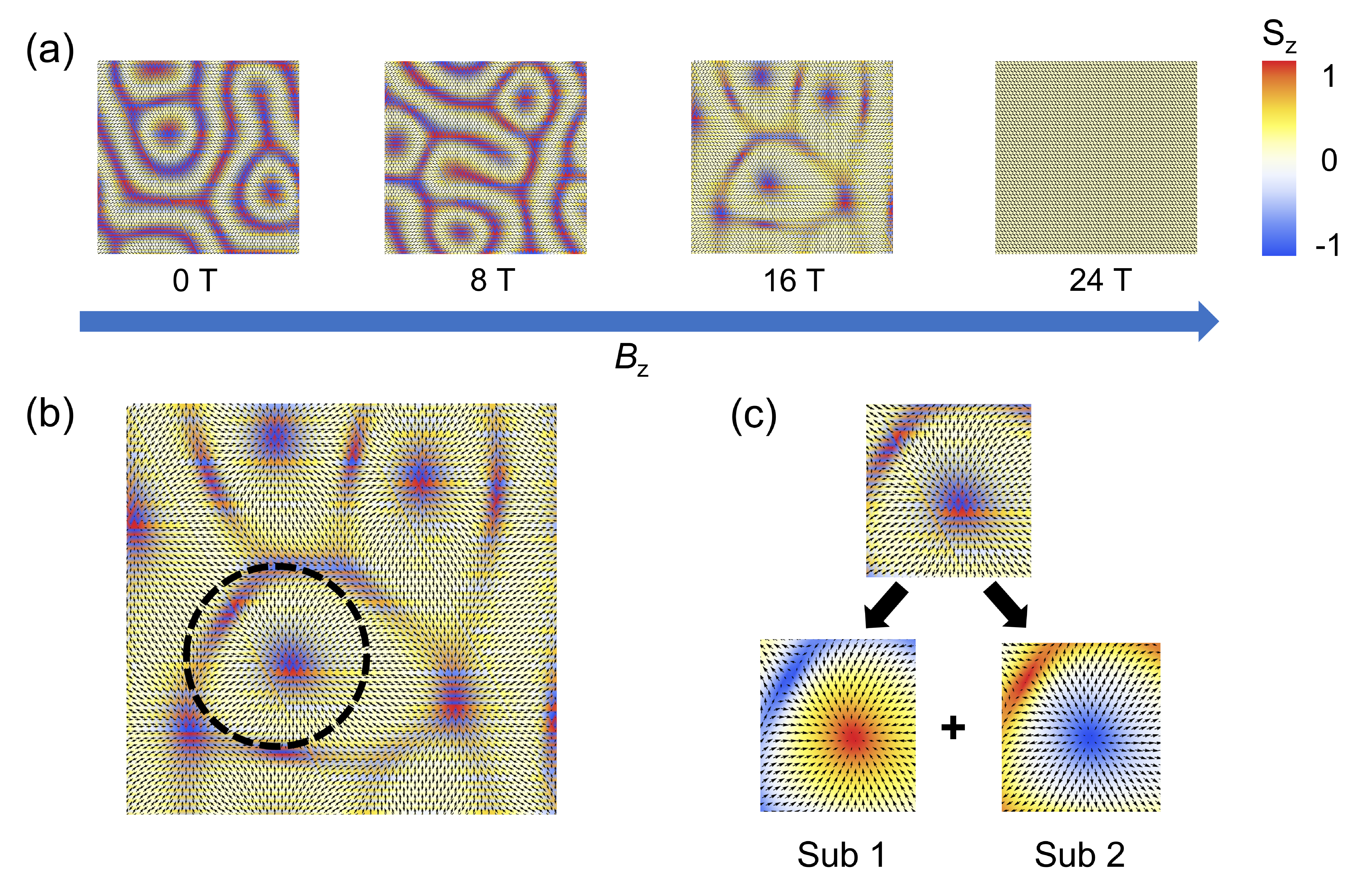}
\caption{\label{6} Field-dependent spin textures. (a) Macroscopic magnetic configurations under distinct magnetic field levels; (b) The close view of magnetic texture under a 16 T external field. Isolated AFM merons are circled. (c) Schematic illustration of isolated AFM merons, which are composed of FM merons on each magnetic sublattice.  }
\end{figure*}

In addition to the DMI effect, the 2NN AFM exchange interaction ($J_{2}$), which competes with the 1NN AFM exchange interaction ($J_{1}$), also plays a significant role in the energy gain of the AFM SS state. To investigate the specific physical impacts of $J_{2}$, we systematically adjusted its value while keeping all other magnetic parameters constant. Through MC simulations with incrementally increased $J_{2}$ values, we observed three interesting phenomena: (1) the period of the lowest-energy AFM SS state decreases monotonically, as shown in Fig. \ref{5}(a); (2) the energy gain relative to the collinear N$\Acute{\textup{e}}$el-type AFM state increases, as illustrated in Fig. \ref{5}(b); and (3) the macroscopic LD state becomes narrower, and the size of isolated skyrmions diminishes, as depicted in Fig. \ref{5}(c). These findings indicate that the exchange frustration between $J_{1}$ and $J_{2}$ not only modulates the size of the AFM chiral spin states but also contributes to stabilizing the AFM SSs and skyrmions against the  N$\Acute{\textup{e}}$el-type AFM background.

\subsection{Field-dependent magnetic configurations}
Generally, when an external magnetic field is applied, the FM SS undergoes a transition to a skyrmion lattice before eventually reaching a field-polarized FM state \cite{PhysRevB.98.060413,Janus3,Janus1}. A similar scenario occurs for the AFM SS state on a geometrically frustrated triangular lattice \cite{PhysRevB.92.214439,dou2022theoretical}. However, less is known about the field-dependent behavior of AFM SS and the metastable LD state in realistic systems with nonfrustrated hexagonal lattices. Hence, it is necessary to qualitatively investigate the evolution of the AFM LD state under a magnetic field (labeled $B_{z}$) in our system. For this purpose, we conduct MC simulations at different field strengths by introducing a Zeeman term, $H_{z} = g\mu _{B}S_{z}B_{z}$, where $g$ is the Land\'e factor and $\mu _{B}$ represents the Bohr magneton. The resulting field-dependent macroscopic spin configuration is displayed in Fig. \ref{6}(a). When $B_{z}$ is weak, both the AFM LDs and the skyrmions exhibit minimal responses to the external magnetic field. This indicates the robustness of AFM topological defects against certain external fields, a trait advantageous for field-immunity spintronic devices. However, as $B_{z}$ increases further, domains with out-of-plane magnetization contract, while domain walls with in-plane spin orientations expand. At a $B_{z}$ value of 16 T, the cores of topological spin textures are surrounded by  domains in which the spins orient in the x-y plane, forming AFM merons, as shown in Figs. \ref{6}(b) and \ref{6}(c). These merons consist of two sets of FM merons on each magnetic sublattice. To the best of our knowledge, the skyrmion-to-meron transition induced by a magnetic field is a distinctive feature of AFM systems, unlike the meron-to-skyrmion transition observed in the FM system \cite{Meron2}. This finding enhances our comprehension of field manipulations of topological spin textures, potentially advancing the development of related AFM spintronic devices. When $B_{z}$ reaches 24 T, all noncollinear local spin textures are suppressed, leading to the emergence of a canted AFM state. In this state, spins undergo slight canting along the z direction due to the presence of the out-of-plane magnetic field (see Fig. S5 in SM \cite{SM}). These findings suggest the feasibility of manipulating AFM topological spin textures with a magnetic field.

\section{CONCLUSIONS}
In conclusion, using monolayer Janus CrSi(Te,Se)$_{3}$ as a representative system, we demonstrate the spontaneous emergence of intrinsic AFM skymions in 2D Janus magnets. The interplay between the strain-induced AFM exchange interactions and substantial DMI in the properly compressed monolayer CrSi(Te,Se)$_{3}$ results in the formation of intrinsic AFM skyrmions, which are made up of AFM SS states.
Furthermore, the application of an external magnetic field induces the emergence of AFM merons as well as a canted AFM state. Our findings propose a feasible pathway for realizing intrinsic AFM topological spin textures in practical 2D systems, offering significant potential for future applications in nanotechnology devices within the realm of AFM spintronics. 

\section{ACKNOWLEDGMENTS}
The authors thank Prof. Wenhui Duan at Tsinghua University and Prof. Changsong Xu at Fudan University for helpful discussions. 
\bibliographystyle{apsrev4-1}

\end{document}